\documentclass[10pt, conference, compsocconf]{IEEEtran}

\usepackage[ruled,vlined]{algorithm2e}
\usepackage{graphicx}
\usepackage{color}
\usepackage{amssymb}
\usepackage{amsmath}
\newtheorem{theorem}{{\bf Theorem}}
 \newtheorem{lemma}{{\bf Lemma}}

\ifCLASSINFOpdf
  
\else
   
\fi

\begin{document}
 
\title{Mutual Visibility by Robots with Persistent Memory}

\author{\IEEEauthorblockN{Subhash Bhagat}
\IEEEauthorblockA{Advanced Computing and Microelectronics Unit\\
 Indian Statistical Institute \\
  Kolkata, India\\
 sbhagat\_r@isical.ac.in}
\and
\IEEEauthorblockN{Krishnendu Mukhopadhyaya}
\IEEEauthorblockA{Advanced Computing and Microelectronics Unit \\
 Indian Statistical Institute\\
  Kolkata, India\\
 krishnendu@isical.ac.in}
}
 
\maketitle

\begin{abstract}
 This paper addresses the {\it mutual visibility} problem for a set of semi-synchronous, opaque robots occupying distinct positions in the  Euclidean plane. Since robots are opaque, if three robots lie on a line, the middle robot obstructs the visions of the two other robots. The mutual visibility problem asks the robots to coordinate their movements to form a configuration, within finite time and without collision, in which no three robots are collinear. Robots are endowed with a constant bits of persistent memory. In this work, we consider the  FS{\scriptsize TATE} computational model in which the persistent memory is used by the robots only to remember their previous internal states. 
Except from this persistent memory, robots are oblivious i.e., they do not carry forward any other information from their previous computational cycles.  
 The paper presents a distributed algorithm  to solve the mutual visibility problem for a set of semi-synchronous robots using only 1 bit of persistent memory. 
 The proposed algorithm does not impose any other restriction on the capability of the robots and guarantees collision-free movements for the robots.

\end{abstract}


\IEEEpeerreviewmaketitle

 \section{Introduction}
 A {\it swarm} of robots is a multi-robot system consisting of autonomous, homogeneous, small mobile
robots which are capable of carrying out some task in a cooperative environment. The robots are modelled as points on the two-dimensional plane in which they can move freely. The robots do not have individual identities i.e., they are indistinguishable by their appearances. Robots are homogeneous i.e., they have same capabilities. They do not have a global coordinate system, 
each robot has its own local coordinate system. The robots sense the positions of the other robots w.r.t. their local coordinate systems. 
Each robot executes the computational cycles consisting of three phases {\it Look-Compute-Move}. In {\it Look} phase, a robot takes the snapshot of its surroundings 
and maps the locations of the other robots w.r.t. its local coordinate system. In {\it Compute} phase, a robot uses the information gathered in the Look state to compute a destination point to move to. In {\it Move} phase, it moves to its computed destination point. Majority of works in the literature assume that the robots are oblivious i.e., they do not remember any data of their  previous computational cycles. All the robots execute same algorithm.

 Three main computational models are studied in the literature. In the asynchronous model (ASYNC or CORDA) \cite{prencipe01}, the scheduling of activities of the robots are unpredictable and independent of each other. 
 However, the duration of each computational cycle is finite. In semi-synchronous model (SSYNC) \cite{Ichuri1993}, time is discretized into several rounds. In each round, a subset of robots is allowed to execute their computational cycles simultaneously. The movement of the robots are instantaneous i.e., a robot is not observed by the other robots while in motion. The fully-synchronous model (FSYNC) requires all the robots to execute their cycles in a single round. We assume a fair scheduler which activates
each robot infinitely often \cite{defago}.

In terms of capabilities of the robots, different assumptions are made to solve the problems. {\it Multiplicity detection} allows a robot to identify multiple 
occurrences of  robots at a single point. Common $chirality$ helps the robots to agree on a common orientation i.e., agreement on common clockwise direction. 
{\it Rigid motion} permits the robots to reach their destinations without halting in between. In persistent memory model, robots are endowed with constant amount of persistent memory (the robots are otherwise oblivious) \cite{das2012}. This persistent memory can be used in three different ways: (i) the robots can set limited communications between themselves using {\it visible lights} which can assume a constant number of predefined colors to represent their different states and also to retain some constant amount of information about their previous states or (ii) only to remember information about their last states (FS{\scriptsize TATE} model) or (iii) the robots can use visible lights only to communicate with other robots in the system and they do not remember the colors of the lights of their last computational cycle (FC{\scriptsize OMM} model) \cite{FlSVY13} . Thus, the persistent memory can be used for communication or for internal memory or for both. In this work, robots use persistent memory only for 
internal memory. {\it Unlimited visibility range} allows  a robot to sense other robots from any distance. {\it Transparency} of the robots provides a obstruction free vision for the 
robots. There can be some agreement on the direction and orientation of the local coordinate axes of the robots. 
 
 The algorithms are designed to coordinate the motion of the robots to solve a variety of problems. Fundamental geometric problems like $gathering$, {\it circle formation}, $flocking$ etc. have been studied extensively in the literature \cite{FlPS12}. Recently some researchers have taken up the problem of 
 {\it mutual visibility} \cite{AFPSV14,sruti2016,log-ipdps,bhagat2016}. The {\it mutual visibility} problem is defined as follows: for a set of robots initially occupying distinct positions in the two dimensional plane, the mutual visibility problem asks the robots to form a configuration, within finite time and without collision, in which no three robots are collinear.

\subsection{Earlier works}

  Most of the investigations on different geometric pattern formation problems assume that the robots are transparent. {\it Obstructed visibility} has been considered for $fat$ robots (robots represented as unit discs) \cite{AgGM13,BoKF12,CzGP09} as well as for the point robots \cite{AOSY99,bhagat2015,CoP08}. Explicit
  communication among the robots using externally visible lights introduced by David Peleg \cite{Peleg2005}. Combining this limited form of communication and 
  memory  with the traditional models, different problems have been solved by many researchers \cite{DasFPSY12,DasFPSY14,EfP07,FlSVY13,Vi13}.  
    Di Luna et. al. \cite{AFPSV14} presented  a distributed algorithm to solve the mutual visibility problem for a set of oblivious, semi-synchronous robots. Sharma et. al. \cite{sharmaCCCG15} analysed and modified the round complexities of the mutual visibility algorithms presented in \cite{AFPSV14} under fully synchronous model.
   Di Luna  et al. \cite{sruti2016} were the first to study the mutual visibility problem  in the {\it light} model. They  solve the problem for the semi-synchronous robots with 3 colors and for asynchronous robots with 3 colors under one axis agreement (in \cite{AFGSV14} authors claimed a solution of the mutual visibility problem for the asynchronous robots with 10 colors. However later in \cite{sruti2016}, they modified their claim and presented a solution for the asynchronous robots with 3 colors under one axis agreement).  Sharma et. al. \cite{sharmaALGo15} proved that the problem is solvable using only 2 colors for the semi-synchronous robots and for the asynchronous robots under one axis agreement.  Vaidyanathan et. al. \cite{log-ipdps} proposed a distributed algorithm for  fully-synchronous robots using 12 colors. The algorithm runs in $O(\log(n))$ rounds for $n\ge4$ robots. The only solution to the mutual visibility problem for asynchronous oblivious robots has been proposed in \cite{bhagat2016} 
under the assumption that  the robots  have an agreement in one coordinate axis and they have knowledge of total number of robots in the system. Thus, all the existing solutions for the mutual visible problem either assume persistent memory for both communication and internal memory purposes or one axis agreement or the knowledge of $n$, total number of robots in the system.
%

\subsection{Our Contribution}
This paper studies the {\it mutual visibility} problem for a set of semi-synchronous robots on the Euclidean plane. A distributed algorithm has been proposed to solve the problem for a set of robots endowed with a constant amount of persistent memory. The proposed algorithm considers FS{\scriptsize TATE} model which does not have communication overhead of FC{\scriptsize OMM} model. The persistent memory is used only to remember information about their previous states. The proposed algorithm does not assume any other extra assumptions like  agreement on the coordinate axes or chirality, knowledge of $n$, rigidity of movements. In spite of these weak assumptions, it is showed that the mutual visibility problem is solvable for a set of semi-synchronous robots using only 1 bit of persistent internal  memory. The contribution of this paper has mainly two folds of significance. First, while all the existing solutions of the mutual visibility problem for semi-synchronous robots have considered  either knowledge of 
$n$ or persistent memory for both communication and internal memory purposes (combination of FS{\scriptsize TATE} and FC{\scriptsize OMM} model), our approach assumes FS{\scriptsize TATE} model without knowledge of $n$ (this makes system easily scalable). Secondly, in all the existing solutions for the mutual visibility problem under persistent memory model, the convex hull of the initial positions of the robots does not remain invariant and the robots move even if the configuration is completely visible to all the robots (robots do not have knowledge of $n$). The solution of this work maintains the convex hull of the initial robot positions if all the robots initially do not lie on a single and if the configuration is completely visible to each robot, the robots do not move. The solution also provides collision free movements for the robots. To the best of our knowledge, this paper is the first attempt to study the mutual visibility problem under FS{\scriptsize TATE} model. 
\section{Model and Notations}
\label{model}

This paper considers a set of $n$ homogeneous, autonomous robots represented by points in the two dimensional Euclidean plane. The robots are capable of moving anywhere they want. 
The robots neither share a global coordinate system nor a common chirality. Each robot has its own local coordinate system; the directions and the orientations
of coordinate axes and the unit distance may be vary . The robots are opaque. However, the visibility range of a robot is unlimited. The robots operate in \emph{look-compute-move} 
cycles repeatedly. The robots are semi-synchronous (SSYNC model). The robots have no knowledge about the total number of robots in the system. 
The movements of the robots are {\it non-rigid} i.e., a robot can be stopped by an adversary before reaching its destination. However, it is assumed that a robot, if it does not reach its destination, must travel a minimum distance $\delta >0$ towards its destination whenever it decides to move. The value of $\delta$ is not known to the robots. The robots do have any explicit communication power. However, each robot has 1 bit of internal persistent memory FS{\scriptsize TATE} model. The 1 bit memory stores information about predefined specific states of the robot. This internal bit does not change automatically and it is persistent. Let $s_i(t)$ be the binary variable which denote the value stored in the internal memory of the robot $r_i$ at time $t\in\mathbb N$. Except for this persistent memory, the robots are oblivious i.e., they do not remember any other data of their previous computational cycles. Initially all the robots occupy distinct locations and they are stationary.

\begin{itemize}
\item \textbf{configurations of the robots:} Let $\mathcal R=\{r_1, r_2,\ldots, r_n\}$ denote the set of $n$ robots. The position of robot $r_i$  at time $t$ is denoted by $r_i(t)$. 
 A configuration of robots, $\mathcal R(t)=\{r_1(t), \ldots, r_n(t)\}$, is the set of positions occupied by the robots at time $t$. $\widetilde{C}$ denotes the set of all such robot configurations. 

  We partition $\widetilde{C}$ into two classes: $\widetilde{C}_L$
 and $\widetilde{C}_{NL}$, where $\widetilde{C}_L$ is the collection of configurations in which all the robots in $\mathcal R$ lie on a straight line and $\widetilde{C}_{NL}$ consists of configurations in which there
 exist at least three non-collinear robot positions occupied by the robots in $\mathcal R$.
 We say that a robot configuration  $\mathcal R(t)$ is in $\textit{general position}$ if no three robot positions in  $\mathcal R(t)$ are collinear.
 By $\widetilde{C}_{GP}$, we denote the set of all configurations of $\mathcal R$ which are in general position. Clearly $\widetilde{C}_{GP} \subset \widetilde{C}_{NL}$.
 
\item \textbf{Measurement of angles:} By an angle between two line segments, if not stated otherwise, we mean the angle which is less than or equal to $\pi$.
\item \textbf{Vision of a robot:} If three robots $r_i, r_j$ and $r_k$ are collinear with $r_j$ lying in
between $r_i$ and $r_k$, then $r_i$ and $r_k$ are not visible to each other. We define the vision,  ${\cal V}(r_i(t))$, of  robot $r_i$ at time $t$ to be
the set of robot positions visible to $r_i$ (excluding $r_i$). The {\it visibility polygon} of $r_i$ at time $t$, denoted by $STR(r_i(t))$, is defined as follows: 
sort the points in ${\cal V}(r_i(t))$ angularly in anti clockwise direction w.r.t. $r_i(t)$ starting from any robot position in ${\cal V}(r_i(t))$. Then connect 
them in that order to generate the polygon $STR(r_i(t))$ (Figure \ref{Visionofarobot}). 
  \begin{figure}[h]
      \centering
     \includegraphics[scale = .75]{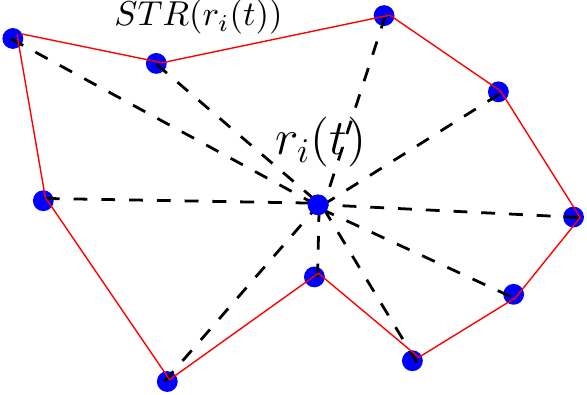}
     \caption{An example of visibility polygon}
     \label{Visionofarobot}
 \end{figure}

 \item  A straight line $\mathcal L$ is called a {\it line of collinearity} if it contains more than two distinct robot positions. A robot occupying a position on $\mathcal L$ is termed a {\it collinear} robot.  For a robot $r_i$, let $\mathcal B_i(t)$ denote the set of all lines of collinearity on which $r_i$ is a collinear robot at time $t\in\mathbb N$. Consider a line of collinearity $\mathcal L$ at time $t$. A robot $r_i$ on $\mathcal L$ is called an \textit{non-terminal} robot if $r_i(t)$ is a point in between two other robot positions on $\mathcal L$. A robot which is not a non-terminal robot is called a \textit{terminal} robot. Let $r_i$ be a non-terminal robot on a line of collinearity $\mathcal L$. The point $r_i(t)$ is called a {\it junction robot position} if there is another line of collinearity $\mathcal L_2$ such that $r_i(t)$ lies at the intersection point between $\mathcal L_1$ and $\mathcal L_2$.

 \item By $\overline{pq}$, we denote the closed line segment joining two points $p$ and $q$, including
the end points $p$ and $q$. Let $(p, q)$ denote the open line segment joining the points
$p$ and $q$, excluding the two end points $p$ and $q$. Let $|\overline{pq}|$ denote the length of $\overline{pq}$.

\item $\boldsymbol{d^k_{ij}(t)}$: Let $\mathcal L_{ij}(t)$ denote the straight line joining $r_i(t)$ and $r_j(t)$.  
The perpendicular distance of the line $\mathcal L_{ij}(t)$ from the point $r_k(t)$ is denoted by $d^k_{ij}(t)$.
\item $\boldsymbol{D_i(t)}$:  $D_i(t)$ is the minimum distance of any two robot positions
 in $\{r_i(t), {\cal V}_i(t)\}$. \\


 \end{itemize}

\section{Algorithm}
\label{algo}
 
The outline of our algorithm is as follows. Consider an initial configuration $\mathcal R(t_0)$ of  robots. If $\mathcal R(t_0)$ contains no non-terminal robot, then $\mathcal R(t_0) \in \widetilde{C}_{GP}$ i.e., all the robots in the system are visible to each other. On the contrary, if $\mathcal R(t_0)$ contains at least one non-terminal robot, then there are at least two robots which are not visible to each other. In this scenario, to achieve complete visibility, robots have to coordinate their movements in such a way that within finite time,  they achieve complete visibility. Regarding the movements of the robots, following three things have to be decided: (i) which robots should move, terminal or non-terminal or both (ii) how much they should move and (iii) the directions of their movements. In our approach to develop a solution for the mutual visibility problem, we choose non-terminal robots for movements until there is no non-terminal robot in the system. The new destination points of the robots are 
computed in such way that (i) they do not create new collinearities  by moving to the new positions and (ii) the total number of collinear robots in the system should decrease within finite number of movements. The algorithm terminates when system contains no   non-terminal robot. A robot can easily determine whether it is a terminal robot or non-terminal robot. A terminal robot does nothing. Before describing the algorithm in details, consider the following cases:\\

\begin{figure}[h]
      \centering
     \includegraphics[scale = .5]{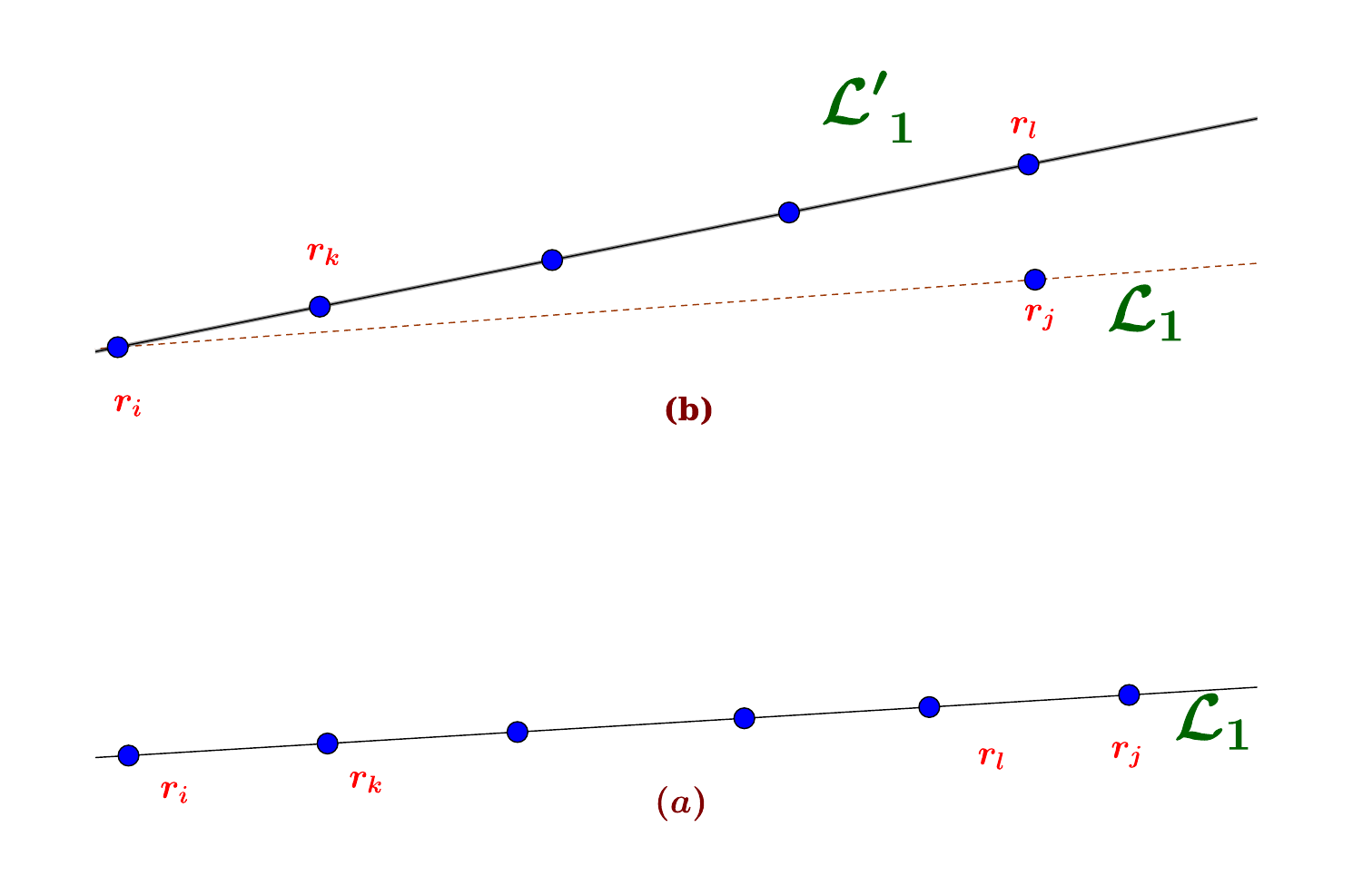}
     \caption{An illustration of case-1 in which non-terminal robots become terminal}
     \label{fig-1}
 \end{figure}

{\bf Case 1:} Let us consider a line of collinearity $\mathcal L_1$. Let $r_i$ and $r_j$ be the two end robots on $\mathcal L_1$ and both of them are terminal robots (Figure \ref{fig-1}(a)). Suppose the robots are activated according to a semi-synchronous scheduler and new destination points of the robots are computed in such a way that no three non-collinear robots in a particular round become collinear in any of 
the succeeding rounds (in the following section, we describe how to compute such points). Suppose only the non-terminal robots on $\mathcal L_1$  move along directions not coincident with $\mathcal L_1$ and all of them move together. Let $r_k$ and $r_l$ be the nearest robots of $r_i$ and $r_j$ respectively on $\mathcal L_1$. After the movements of the non-terminal  robots on $\mathcal L_1$, at least one of the robots among  $r_k$ and $r_l$  becomes terminal. For example, in figure \ref{fig-1}(b), $r_l$ becomes terminal on line $\mathcal L'_1$ (robots may move in opposite sides of $\mathcal L_1$). If the non-terminal robots on $\mathcal L'_1$ move again, at least one of the non-terminal robots on  $\mathcal L'_1$ becomes terminal. In this way, within finite number of movements, all the initially non-terminal robots on $\mathcal L_1$ become terminal and visible to each other. Thus, if a line of collinearity contains two terminal robots and only the non-terminal robots move, in each round at least one non-
terminal robot on this line becomes terminal.\\

\begin{figure}[h]
      \centering
     \includegraphics[scale =.1]{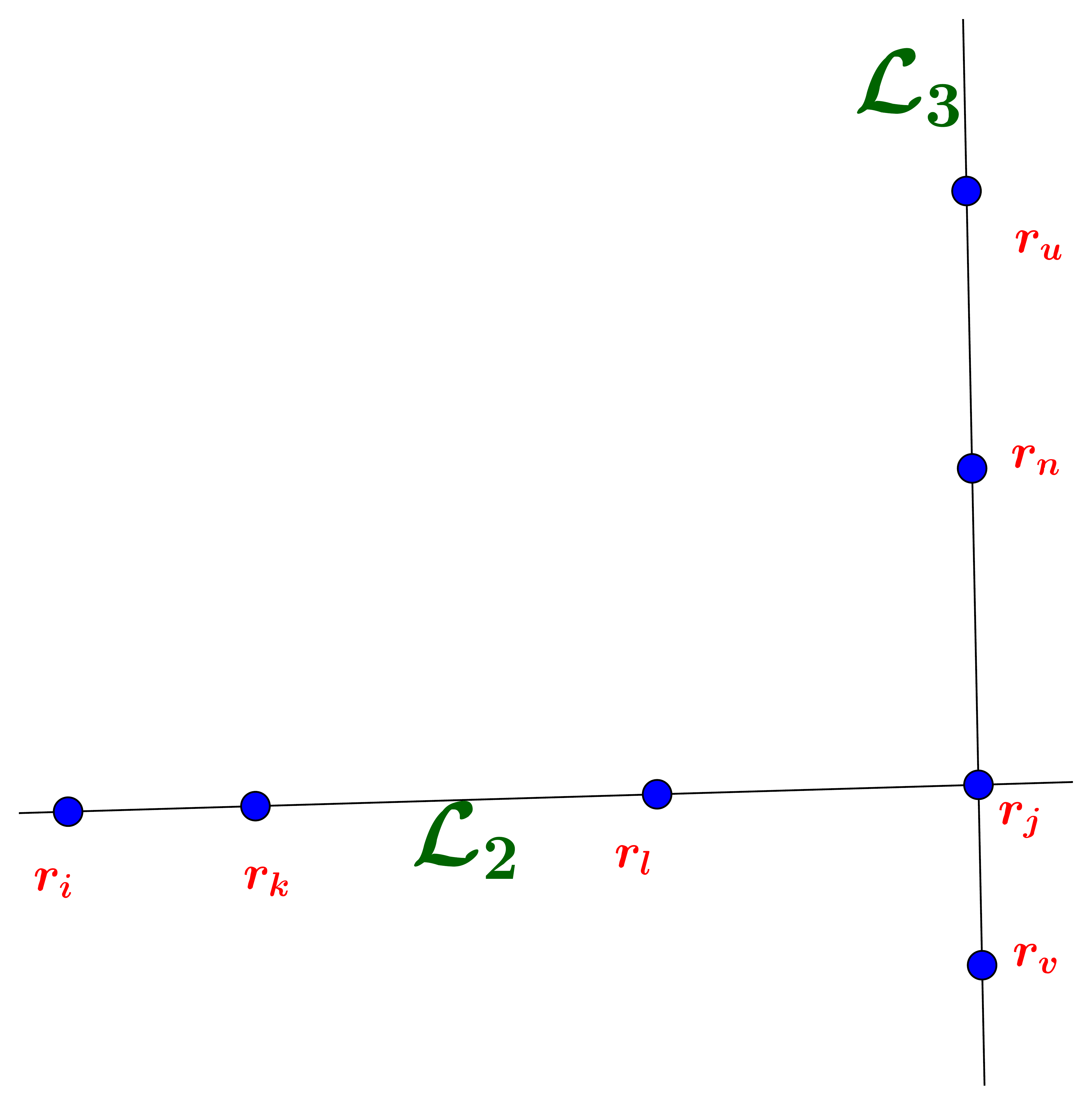}
     \caption{An illustration of case-2 where the line $\mathcal L_2$ contains a junction robot position}
     \label{fig-2}
 \end{figure}

{\bf Case 2:} Let $\mathcal L_2$ be a line of collinearity such that at least one of the two end robot positions on this line is non-terminal position. Consider the case when $\mathcal L_2$ contains exactly one terminal robot, say $r_i$. Let $r_j$ be the robot which occupies the other end robot position on $\mathcal L_2$. Let $r_j$ be a non-terminal robot on a line $\mathcal L_3$ (there may be multiple such lines) (Figure \ref{fig-2}). 

\begin{figure}[h]
      \centering
     \includegraphics[scale = .3]{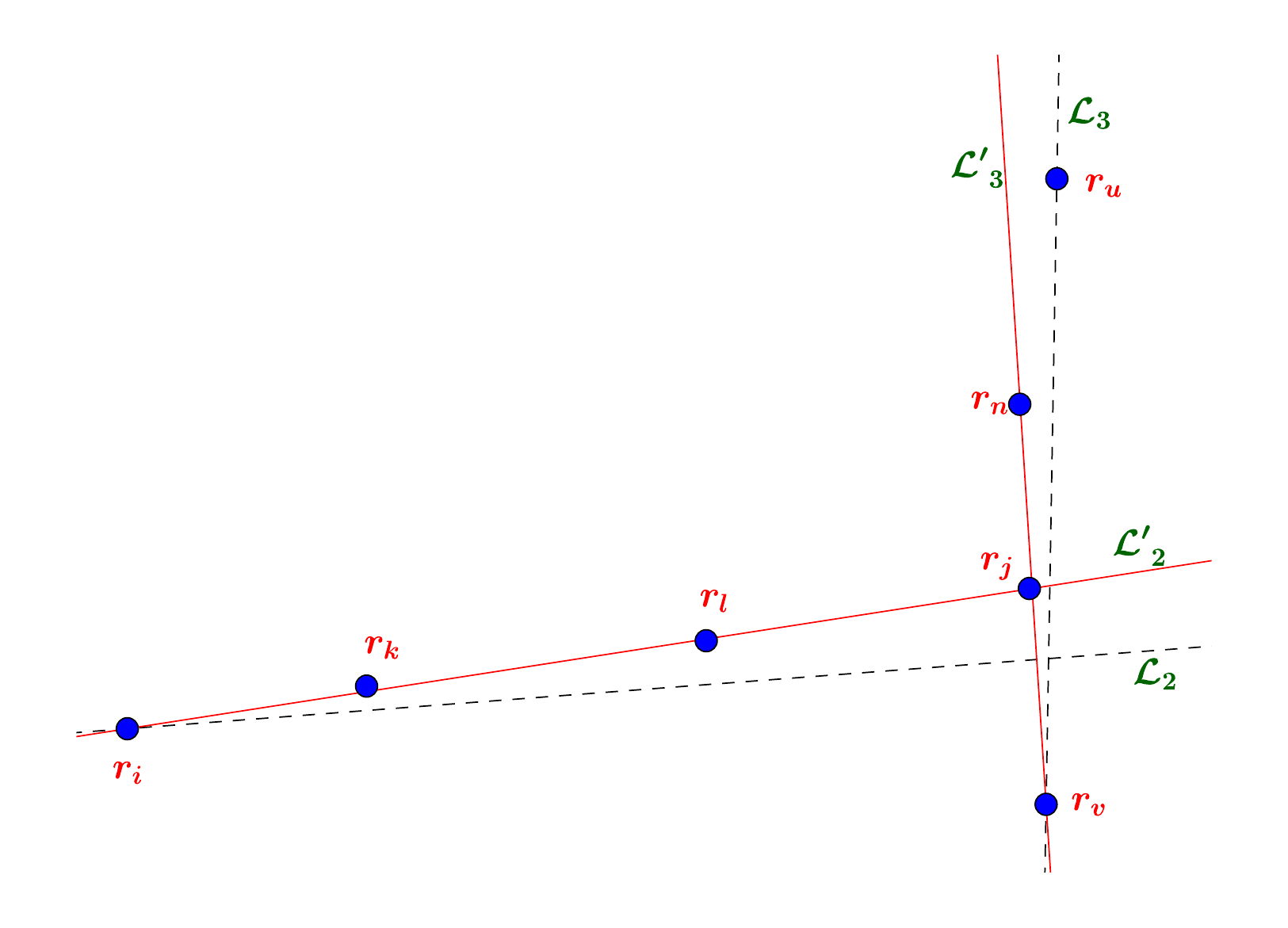}
  \caption{An illustration of case-2 when no non-terminal robot on the line $\mathcal L_2$ becomes terminal}
     \label{fig-3}
 \end{figure}

Suppose all non-terminal robots on $\mathcal L_2$ and $\mathcal L_3$ move, as the same way as in case-1. It may happen that all the non-terminal robots on $\mathcal L_2$ remain collinear with all the robots on  $\mathcal L_2$ i.e.,  the line $\mathcal L_2$ is just shifted to new position with all the robots on it (the line $\mathcal L_2$ is rotated about the point $r_i(t)$) (Figure \ref{fig-3}). Thus, from the line $\mathcal L_2$, no non-terminal robot becomes terminal.

\begin{figure}[h]
      \centering
     \includegraphics[scale = .3]{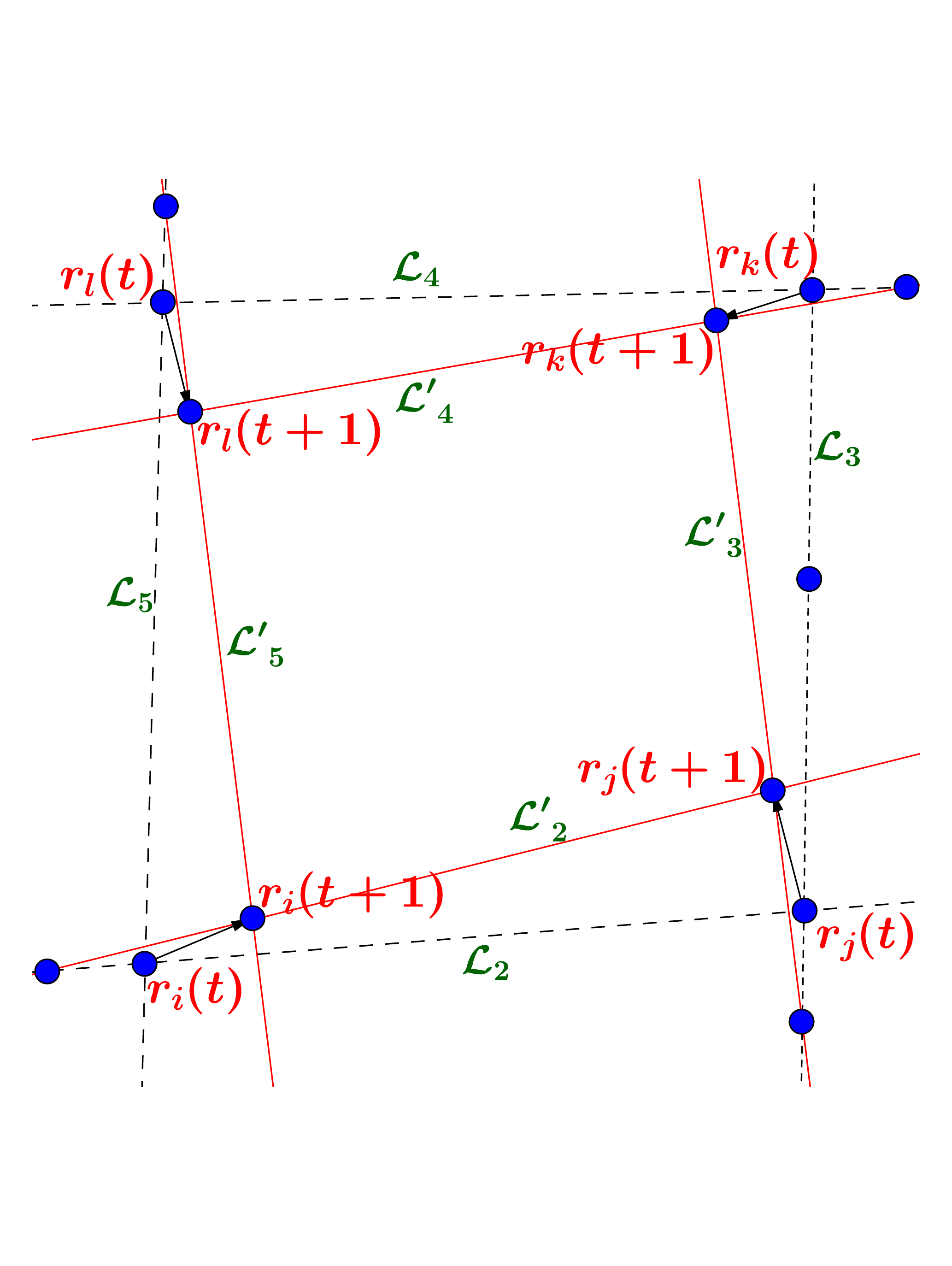}
     \vspace*{-.2in}   
     \caption{An illustration when the non-terminal robots in a {\it cycle} remain non-terminal even after their movements}
     \label{fig-41}
 \end{figure}

\begin{figure}[h]
      \centering
     \includegraphics[scale = .25]{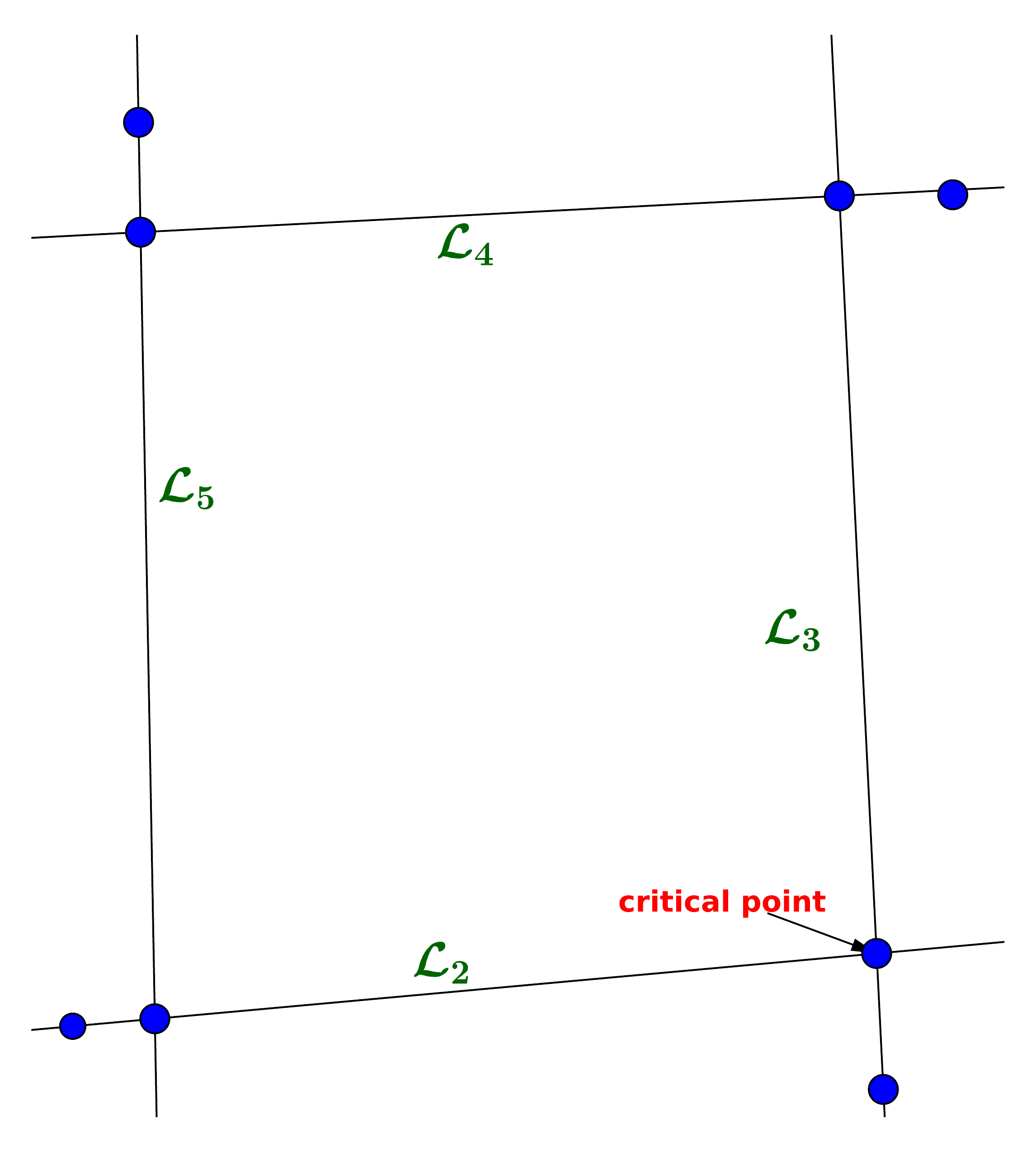}
     \caption{An example of a cycle for $k=4$}
     \label{fig-4}
 \end{figure}

 If the line $\mathcal L_3$ contains two terminal robot and exactly one junction robot position, then by case-1, the movements of the robots creates at least one terminal robot. However, if $\mathcal L_3$ contains at most one terminal robot, by foregoing arguments, all the non-terminal robots on $\mathcal L_3$ may remain non-terminal just like the case of $\mathcal L_2$. In this way, we can get  cyclic dependencies between the lines of collinearity such that the movements of the non-terminal robots may not create new terminal robots within finite number of movements (Figure \ref{fig-41}). Let us formally define this 
cyclic dependency. 
Let $\mathcal S=\{L_2, L_3, \ldots, L_k\}$ be a sequence of lines of 
collinearity. We say that this sequence of lines form a {\it cycle} if each of the lines in this sequence contains more than one junction robot positions and one junction robot position on the line $\mathcal L_m$ lies on the line $\mathcal L_{m+1}$ and one junction robot position lies on the line $\mathcal L_{m-1}$ where $m\ge 2$, $\mathcal L_{1}$ is $\mathcal L_{k}$ and $\mathcal L_{k+1}$ is $\mathcal L_2$ (Figure \ref{fig-4}). The robot positions at the intersection point between two lines in a cycle are called  {\it critical} points.

\begin{figure}[h]
      \centering
     \includegraphics[scale = .2]{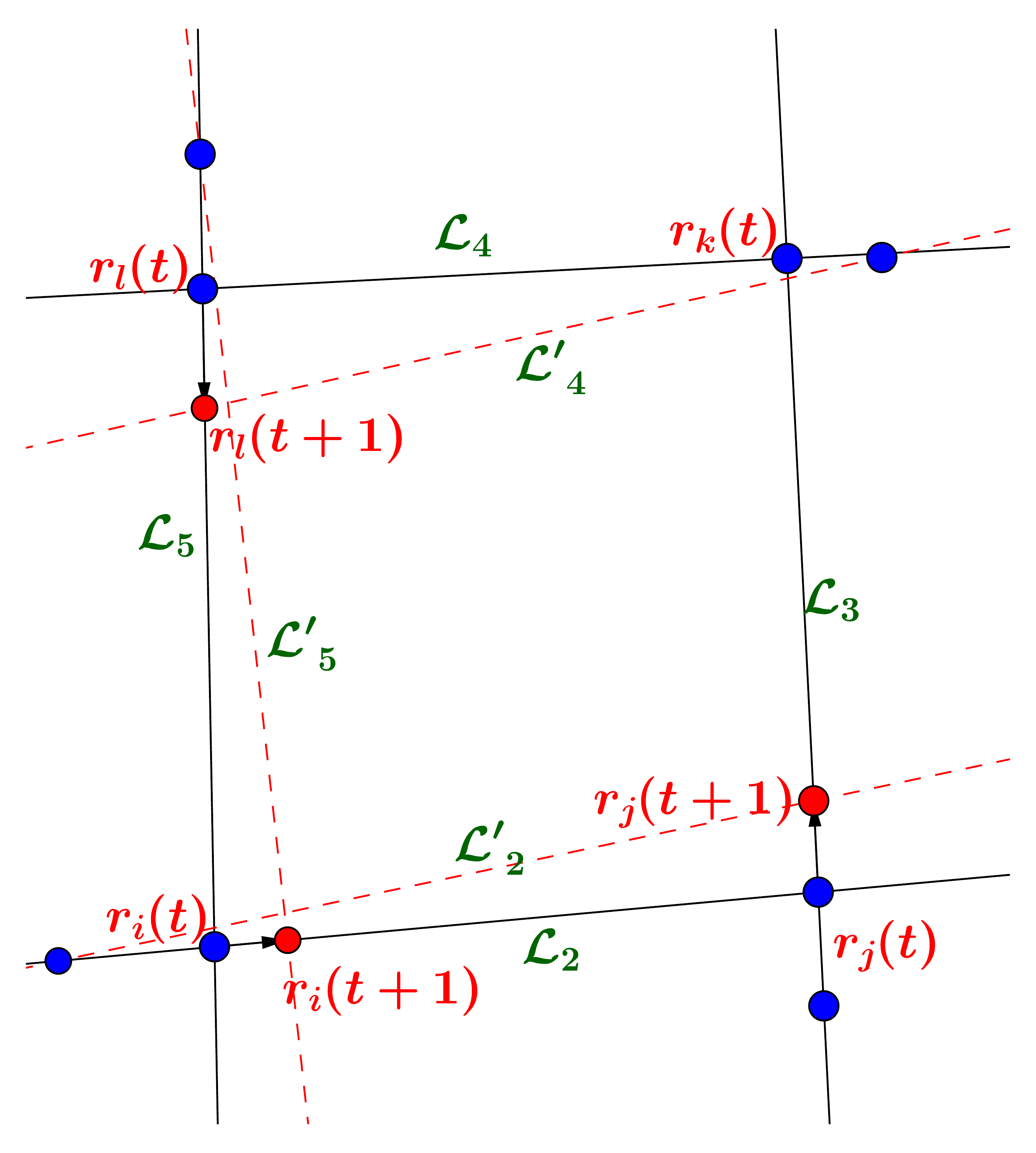}
     \caption{An illustration when the non-terminal robots in a {\it cycle} become terminal due to  type-1 movements}
     \label{fig-42}
 \end{figure}

The question is how to break the cyclic dependency among the lines of collinearity? One of the ways is as follows: if the non-terminal robots move along the their corresponding lines of collinearity, then this cyclic dependency can be broken within one round (Figure \ref{fig-42} ).

The strategy to break collinearity in case-2 does work for case-1 and vice versa. To break all collinearities by moving the non-terminal robots, we need to combine both the strategies applied in case-1 and case-2. Since in general robots are oblivious, we can not combine both the strategies stated in case-1 and case-2.  In our model, robots are endowed with 1 bit of persistent memory and   this memory can be used to get ride of the difficulties in combining the two strategies. Robots use their internal memories to remember the information about  two types of  movements as stated in case-1 and case-2. Robots use 0 and 1 in their persistent memory for this purpose. Initially all robots have 0 in their respective 1 bit of persistent memory. If the internal bit is 0, a robot moves not along any line of collinearity and this move is called a {\it type-0} move. If internal bit is 1, a robot moves along a line of collinearity and this move is called a {\it type-1} move.   

The above tow cases illustrate the need for considering 1 bit of internal memory. Now, we describe our algorithm in details. The the computations of the destination points depend upon whether the initial configuration $\mathcal R(t_0)$ is in $\widetilde{C}_{L}$ or in $\widetilde{C}_{NL}$. If the initial configuration is in $\widetilde{C}_{L}$, movement of any robot converts this into a configuration in $\widetilde{C}_{NL}$. The complete description of our strategies are as follows.

 \subsection{Different types of movements}  Type-0 and Type-1 moves, as defined above. 
 \subsection{States of a robot} A robot uses its persistent 1 bit memory to remember information about its last movement. 
 Initially all robots have 0 in their persistent memory.
 \begin{itemize}
  \item If a robot is terminal and its internal bit is 0, it  is a terminal robot since the initial configuration.
  \item If a robot is terminal and its internal bit is 1, it was a non-terminal robot in the initial configuration and has become terminal during the execution of the algorithm.
  \item If a robot is non-terminal and its internal bit is 0, it is a non-terminal robot since the initial configuration and either it has made no move or has made a type-1 move.
    \item If a robot is non-terminal and its internal bit is 1, it is a non-terminal robot since the initial configuration and   it has made a type-0 move.
 \end{itemize}

\subsection{Eligible robots for movements} The non-terminal robots are eligible for movements. The terminal robots does nothing.

\subsection{ Computation of destination point}

Let $r_i$ be an arbitrary non-terminal robot at time $t\ge t_0$. To find the new position of $r_i$, we first decide on the direction of movement and then the amount of displacement along the this direction. 
While computing the new destination point of $r_i$, two things should be taken care of. One is that the new position of $r_i$ should not block the visibility of the other robots and the second one is that
the motions of the robots should be collision free. Depending upon the current configuration $\mathcal R(t)$, the destination point for $r_i$ is computed as follows.

\begin{itemize}
 \item \textbf{Case-1:} $\mathbf{\mathcal R(t) \in \widetilde{C}_{NL}}$\\
 Consider the set  of angles  $\Gamma(r_i(t))$ defined as follows:\\ 
 
 $\Gamma_i(t)=\{\angle{r_jr_ir_k}:$ $r_j,r_k$ are two consecutive vertices on $STR(r_i(t))\}$\\
 
  \begin{itemize}
   \item \textbf{The direction of movement:}  Let $\alpha_i(t)$ denote the angle in
 $\Gamma_i(t)$ having the maximum value if the maximum value is less than $\pi$, otherwise the $2^{nd}$ maximum value (tie, if any, is broken arbitrarily). 
  The bisector of $\alpha_i(t)$ is denoted  by $Bisec_i(t)$. It is a ray from $r_i(t)$. If persistent bit is 0,  $r_i$ makes a type-0 move and its the direction of movement is along $Bisec_i(t)$. Before starting its movement, $r_i$ changes its persistent bit to 1. It may be noted that any other suitable direction for type-0 move would work fine for robot $r_i$. If persistent bit is 1, $r_i$ makes a type-1 move. $r_i$ randomly chooses a line of collinearity from $\mathcal B_i(t)$ and moves along this line. Before starting a type-1 move, $r_i$ changes its persistent bit to 0.  
\item \textbf {The amount of displacement:} 
  
Let $d_i(t)= minimum\{d^k_{ij}(t), d^j_{ik}(t), d^i_{jk}(t): \forall r_j, r_k \in {\cal V}_i(t))\}$.  
 The amount of displacement of $r_i$ at time $t$ is denoted by $\sigma_i(t)$  and it is defined as follows,
 
\begin{equation*}
\sigma_i(t)=  \frac{U}{3^{4v_i(t)}} 
\end{equation*}
Where {\it U=minimum}$\{d_i(t), D_i(t)\}$   and  $v_i(t)=|{\cal V}_i(t))|$.

Three non-collinear robots become collinear when the triangle formed by these their positions diminishes to a line. The amount $\sigma_i(t)$ is chosen to be a small fraction of $d^k_{ij}(t)$ for all  $r_j(t), r_k(t)\in {\cal V}_i(t))$ in order to guarantee that no new collinearity is generated during the movements of the robots. 
Other suitable values will also work. 
\item \textbf {The destination point:} Let $\hat r_i(t)$ be the point on $Bisec_i(t)$ at distance $\sigma_i(t)$ from $r_i(t)$ if $s_i(t)=0$. Otherwise, $\hat r_i(t)$ is a point on a line $\mathcal L\in\mathcal B_i(t)$ at distance $\sigma_i(t)$ from $r_i(t)$ (choose randomly any one of the two directions along $\mathcal L$).  The destination point of $r_i(t)$ is $\hat r_i(t)$.\\
\end{itemize}

 \item \textbf{Case-2:} $\mathbf{\mathcal R(t) \in \widetilde{C}_{L}}$ \\ 
 There is only one line of collinearity, say $\hat {\mathcal L}$, in the system. Only two robots are terminal. 
 Once one of them moves, the present configuration is converted into a configuration in $\widetilde{C}_{NL}$. 
 \begin{itemize}
  \item \textbf{The direction of movement:} Let $\mathcal L^*$ be the perpendicular line to $\hat {\mathcal L}$ at the point $r_i(t)$. The robot $r_i$ arbitrarily chooses a direction along $\mathcal L^*$ and moves along that direction. Let $\mathcal L^*_d$ denote the direction of movement of $r_i$. Since all robots are collinear, this movement is a type-0 move. Before starting this move, $r_i$ changes its persistent bit to 1. 
  \item \textbf{The amount of displacement:} In this, the amount of displacement $\hat \sigma_i(t)$ is defined as follows:
\begin{equation*} 
\hat \sigma_i(t)=  \frac{D_i(t)}{3^4} 
\end{equation*}
\item {\bf The destination point:} Let $\bar r_i(t)$ be the point on $\mathcal L^*_d$ at the distance $\hat \sigma_i(t)$ from $r_i(t)$. The
destination point of $r_i$ is $\bar r_i(t)$.

 \end{itemize}
 \end{itemize}
 
 \subsection{Termination} A robot terminates the execution of algorithm $MutualVisibility()$ when it finds itself as a terminal robot. Thus, an initially terminal robot terminates just in one round.
 
 
Robots use the algorithm $ComputeDestination()$ to compute its destination point and use algorithm $MutualVisibility()$ to obtain complete visibility.
\begin{algorithm}
\KwIn{$r_i(t)$, $s_i(t)$ and $\mathcal R(t)$.} 
\KwOut{ a destination point}

\eIf{$|{\cal V}_i(t)|> 2$}
{
$d_i(t)\leftarrow minimum\{d^k_{ij}(t), d^j_{ik}(t), d^i_{jk}(t):\hspace*{3cm} \forall r_j, r_k \in {\cal V}_i(t))\}$\;
$D_i(t)\leftarrow minimum\{|\overline{r_j(t)r_k(t)}|: \forall r_j, r_k \in\hspace*{3cm} \{r_i(t), {\cal V}_i(t)\}\}$\;
$U\leftarrow minimum\{d_i(t), D_i(t)\}$\;
$v_i(t)\leftarrow|{\cal V}_i(t))|$\;
$\sigma_i(t)\leftarrow \frac{1}{3^{4v_i(t)}}U$\; 
\eIf{$s_i(t)=0$}
{
$\alpha_i(t)\leftarrow maximum\{\theta_i(t)\in \Gamma_i(t): \theta_i(t)<\pi\}$\;
$Bisec_i(t)\leftarrow$ Bisector of $\alpha_i(t)$\;
$p\leftarrow$ the point on $Bisec_i(t)$ at a distance $\sigma_i(t)$ $\hspace*{.7cm}$ from $r_i(t)$;
}
{$\mathcal L\leftarrow$ an arbitrary line in $\mathcal B_i(t)$\;
$\mathcal L^+\leftarrow$ any one of the two directions along the $\hspace*{1cm}$ line  $\mathcal L$\;
$p\leftarrow$ the point on $\mathcal L^+$ at a distance $\sigma_i(t)$  from $\hspace*{.2in}$ $r_i(t)$\;
}
}
{
$D_i(t)\leftarrow minimum\{|\overline{r_j(t)r_k(t)}|: \forall r_j, r_k \in\hspace*{3cm} \{r_i(t),
{\cal V}_i(t)\}\}$\;
$\hat\sigma_i(t)\leftarrow \frac{1}{3^{4}}D_i(t)$\;
$\hat{\mathcal L}\leftarrow$ the line in $\mathcal B_i(t)$\;
 $\mathcal L^*\leftarrow$ perpendicular line to  $\hat{\mathcal L}$\;
 $\mathcal L^*_d\leftarrow$ any one of the two directions along the $\hspace*{1cm}$ line  $\mathcal L^*$\;
$p\leftarrow$ the point on $\mathcal L^*_d$ at a distance $\hat \sigma_i(t)$  from $\hspace*{.4in}$ $r_i(t)$\;
}
  
return $p$\;
 
\caption{ComputeDestination()}
\end{algorithm}

\begin{algorithm}
 
\KwIn{$\mathcal R(t)$, a configuration of a set robots $\mathcal R$ .}
\KwOut{$\mathcal R(\hat t)$,  in which no three robots are collinear.}

\eIf{terminal}
{do nothing\;}
{
\eIf{$s_i(t)==0$}
{$p=ComputeDestination(r_i(t), s_i(t), \mathcal R(t))$ \;
$s_i(t)=1$\;}
{
$p=ComputeDestination(r_i(t), s_i(t), \mathcal R(t))$ \;
$s_i(t)=0$\;
}
}
 Move towards $p$ along the line segment $\overline{r_i(t)p}$\;

\caption{$MutualVisibility()$}
\end{algorithm}
\subsection{Correctness}
To prove  the correctness of our algorithm, we need to prove the following: (i) three non-collinear robots in a particular round do not become collinear in any of the succeeding rounds (ii) within finite number of rounds at least one non-terminal robot becomes terminal and (iii) movements of the robots  are collision free. If three non-collinear robots become collinear, then the triangle formed by their positions should collapse into either a line or a point. Thus, for arbitrary three non-collinear robots
 $r_i$, $r_j$ and $r_k$, we prove that  none of $d^k_{ij}(t)$, $d^k_{ij}(t)$ and $d^k_{ij}(t)$  becomes zero. Without loss of generality, we prove that $d^k_{ij}(t)$ will never vanish, during the execution of our algorithm. We estimate the maximum decrement in the value of $d^k_{ij}(t)$ in a particular round, due to the movements of the robots. \\
\vspace*{-.5cm}

  \begin{lemma}
\label{non-collinear}
Let $r_i,r_j$ and $r_k$ be three arbitrary robots, which are not collinear at time $t\in \mathbb N$. During the rest of execution of algorithm $MutualVisibility()$, they do not become collinear.
 
\end{lemma}
  
 \textbf{Proof.} Maximum decrement in the value of $d^k_{ij}(t)$ occurs when all the three robots move simultaneously in a round. Thus, we suppose the three robots move at time $t$.  
 Depending upon the positions of the robots, we have the following cases.

  \begin{itemize}
   \item \textbf{Case-1: $\boldsymbol{r_i, r_j}$ and $\boldsymbol r_k$ are mutually visible at $\boldsymbol t_0$}\\
    According to our approach, the displacement of a robot, in a single movement, is bounded above by $\frac{d^k_{ij}(t)}{3^{4}}$ (since $|{\cal V}_i(t)|\ge1$). Since all the three robots move simultaneously in a round, the total decrement in the value of $d^k_{ij}(t)$ is bounded above by  $\frac{3}{3^{4}}d^k_{ij}(t)$. It is easy to see that this bound also holds for all other scheduling of the actions of the robots.  
   Thus, we have,
  \begin{equation}
   d^k_{ij}(t+1)>(1-\frac{3}{3^{4}}) d^k_{ij}(t)
  \end{equation}
 Equation (1) implies that the $\triangle_{ijk}(t)$ does not collapses into a line due to the movements of the robots. Since robots are semi-synchronous and $t$ is arbitrary, these three robots never become collinear during the whole execution of the algorithm.

  \item \textbf{Case-2: $\boldsymbol{r_i, r_j}$ and $\boldsymbol r_k$ are not mutually visible at $\boldsymbol t_0$}\\ We show that the triangle $\triangle_{ijk}(t)$ contains another triangle whose three vertices are mutually visible to each other. By case-1, this contained triangle does not vanish during the movements of the robots and so does $\triangle_{ijk}(t)$.  

  \begin{itemize}
  \item \textbf{Case-2.1: Two pairs of robots are mutually visible}\\ Without loss of generality, suppose that $r_j(t),r_k(t)\in {\cal V}_i(t)$ and $r_k(t)\notin {\cal V}_j(t)$. Then there exist two robots $r_u$ and $r_v$ (not necessarily distinct), 
  closest to $r_j$ and $r_k$ respectively, such that they lie
  on $\mathcal L_{jk}(t)$ (Figure \ref{lem2-1}). If $r_u\in {\cal V}_i(t)$, then $r_i,r_j$ and $r_u$ are mutually visible and the triangle $\triangle_{iju}(t)$ is contained within $\triangle_{ijk}(t)$. If  $r_u\notin {\cal V}_i(t)$, there exists a robot $r_x$ such that $r_x$ lies inside the triangle
   $\triangle_{ijk}(t)$ and $r_x$ is visible to both of $r_i$ and $r_j$. In this case, the triangle $\triangle_{ijx}(t)$ is contained within $\triangle_{ijk}(t)$.  
      \begin{figure}[h]
    \centering
   \includegraphics[scale =.8]{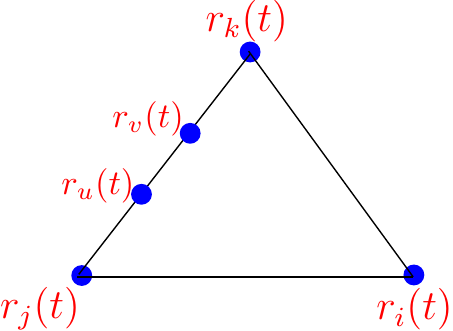}
    \caption{An illustration of Case-2.1 of lemma $\ref{non-terminal}$}
   \label{lem2-1}
  \end{figure}
  
   \item \textbf{Case-2.2: One pair of robots are mutually visible}\\ Without loss of generality, suppose that 
   $r_k(t)\notin {\cal V}_i(t) \cup {\cal V}_j(t)$ and $r_j(t)\in {\cal V}_i(t)$.
   Then there exist (i) two robots $r_{u_1}$ and $r_{v_1}$ (not necessarily distinct), 
  closest to $r_i$ and $r_k$ respectively, such that they lie
  on $\mathcal L_{ik}(t)$ and (ii) two robots $r_{u_2}$ and $r_{v_2}$ (not necessarily distinct), 
  closest to $r_j$ and $r_k$ respectively, such that they lie
  on $\mathcal L_{jk}(t)$ (Figure \ref{lem2-2}).   
 By the same arguments as above,  the triangle $\triangle_{ijx_1}(t)$ is contained within $\triangle_{ijk}(t)$, where $x_1$ is a robot (i) closest to $\mathcal L_{ij}(t)$ (ii) visible to both of $r_i$ and $r_j$ and (iii) lies within or on the triangle $\triangle_{ijk}(t)$ ($x_1$ may be one of $r_{u_1}$ and $r_{u_2}$). 
   \begin{figure}[h]
    \centering
   \includegraphics[scale =.8]{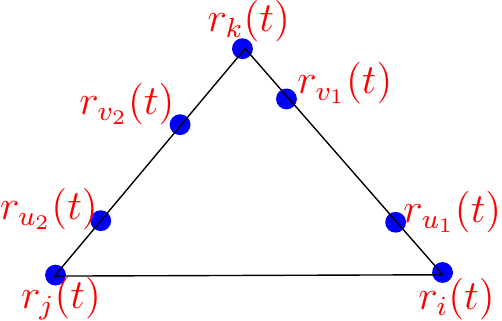}
    \caption{An illustration of Case-2.2 of lemma $\ref{non-terminal}$}
   \label{lem2-2}
  \end{figure}
 
  \item \textbf{Case-2.3: No pair of robots is mutually visible}\\ In this case, 
  $r_i(t)\notin {\cal V}_k(t) \cup {\cal V}_j(t)$ and $r_j(t)\notin {\cal V}_k(t)$. Hence, there exist 
  (i) two robots $r_{u_1}$ and $r_{v_1}$ (not necessarily distinct), closest to $r_i$ and $r_k$ respectively, such that they lie on $\mathcal L_{ik}(t)$ 
  (ii) two robots $r_{u_2}$ and $r_{v_2}$ (not necessarily distinct), closest to $r_i$ and $r_j$ respectively, such that they lie on $\mathcal L_{ij}(t)$ and 
  (iii) two robots $r_{u_3}$ and $r_{v_3}$ (not necessarily distinct), closest to $r_j$ and $r_k$ respectively, such that they lie on $\mathcal L_{jk}(t)$ 
  (Figure \ref{lem2-12}).
   In this case,  the triangle $\triangle_{iu_1u_2}(t)$ is contained within $\triangle_{ijk}(t)$.
    \begin{figure}[h]
    \centering
  \hspace*{2cm} \includegraphics[scale =.8]{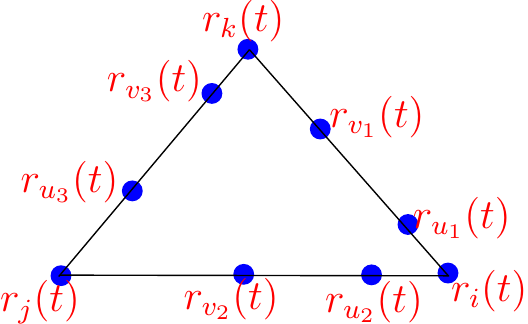}
   \hspace*{1in}\caption{An example of Case-2.3 of lemma $\ref{non-terminal}$}
   \label{lem2-12}
  \end{figure} 
  
  \end{itemize}


  \end{itemize}   
 
Hence the lemma is true.
%
%
%
%

 \begin{lemma}
  \label{non-terminal}
  Let $r_i$ be an initially non-terminal robot. During the execution of algorithm $MutualVisibility()$, $\exists$ a time $t\in \mathbb N$ such that  $r_i$ becomes a terminal robot at time $t$ and it remains terminal for the rest of the execution of the algorithm.

 \end{lemma}
 \textbf{Proof.} 

Let $\mathcal L_1$ be a line of collinearity in $\mathcal B_i(t)$.
 
 \begin{itemize}
  
   \item \textbf{Case-1: $\mathcal L_1$ does not contain a junction robot position}\\
     In this case $l=1$ i.e., $r_i$ is a non-terminal robot on exactly one line. Since both the end robot positions on $L_1$ are terminal, it takes at most $2k-1$ rounds for the non-terminal robots on $\mathcal L_1$ to become terminal, where $k$ is number of non-terminal robots on $\mathcal L_1$.
     \begin{figure}[h]
      \centering
     \includegraphics[scale = .8]{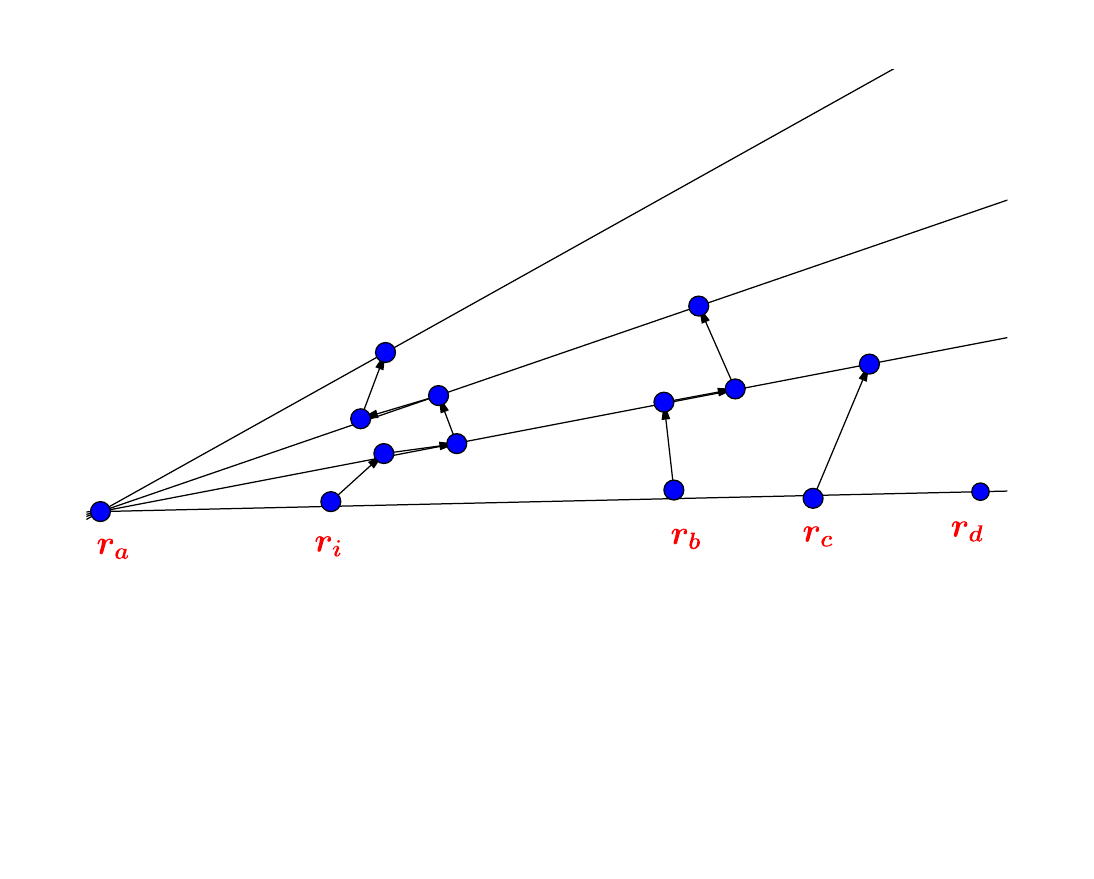}
     \vspace*{-1in}
     \caption{An illustration of case-1 of lemma \ref{non-terminal}}
     \label{fig-5}
 \end{figure}
     \item \textbf{Case-2: $\mathcal L_1$ contains a junction robot position}\\
   We first consider a basic scenario. Let $\mathcal L_1$ contain exactly one junction robot position and $r_k$ be the robot at this position. Let $r_k$ lie exactly on two lines of collinearity and $\mathcal L_2\neq\mathcal L_1$ be the other line of collinearity of $r_k$.
 If $\mathcal L_2$ does not contain any other junction robot position, by case-1.1, within finite round $r_k$ at least occupies one end robot position on either on $\mathcal L_1$ or  $\mathcal L_2$ or on both (if $r_k$ becomes terminal, we are done). 
 Without loss of generality, suppose $r_k$ lies at one end of $\mathcal L_2$ and on $\mathcal L_1$ it is still non-terminal. If $\mathcal L_2$ contains non-terminal robots, they may remain non-terminal due to the movement of $r_k$, until $r_k$ occupies one end robot position on $\mathcal L_1$ i.e., $r_k$ becomes terminal. Once $r_k$ becomes terminal, by case-1.1, the collinearities among the robots initially on $\mathcal L_1$ and $\mathcal L_2$ are broken within finite round and $r_i$ becomes terminal. On the other hand, suppose $\mathcal L_2$ contains another junction robot position, say $r_m$ and $r_k$ and $r_m$ are the only two robots which occupy junction position on $\mathcal L_2$. Let $r_m$ be $\mathcal L_3\neq\mathcal L_2$ be a line of collinearity on which $r_m$ lies.
 If $r_k$ lies exactly on two lines of collinearity ($\mathcal L_2$  and $\mathcal L_3$ ) and $\mathcal L_3$ does not contain a junction robot position, by the same arguments as above,
 within finite round $r_i$ becomes terminal. Suppose $\mathcal L_3$ contains another junction robot position. $\mathcal L_3$  contains exactly two junction robot positions, we are done as above. Otherwise, continuing our arguments as above, we get a sequence $\mathcal S$ of lines of collinearity. Since there are finite number of robots, this sequence either  ends with a line of collinearity $\mathcal L_k$ contain exactly one junction robot position  or it contains a {\it cycle}. If former is true, as above, all the non-terminal robots in this sequence become terminal within finite time. When $\mathcal S$ contains a {\it cycle}, then a type-1 move breaks this {\it cycle}, within finite time. Thus, in this basic scenario within finite number of rounds, $r_i$ becomes terminal.
 
 Now consider the general scenario, in which a line of collinearity may contain more than two junction robot position. Thus, starting from $\mathcal L_1$, we can get many such sequences of lines of collinearity. Let $\widetilde{\mathcal S}$  denotes the set of all these sequence. Since the sequences in $\widetilde{\mathcal S}$ may have common lines, removal of collinearities from one line may depend on the removal of collinearities from another line. If no sequence in $\widetilde{\mathcal S}$ contains a {\it cycle}, then only type-1 movements will break all the collinearities in $\widetilde{\mathcal S}$. Suppose a sequence in $\widetilde{\mathcal S}$ contains a {\it cycle} $\mathcal C$. Let $r_x$ be a robot at a critical robot position on a line $\mathcal L_v$ in $\mathcal C$. If robot $r_x$ makes a type-1 move along $\mathcal L_v$, then  $r_x$ does not remain as a robot at critical position and the cycle $\mathcal C$ is broken. Suppose $r_x$ makes a type-1 move along another line  of collinearity $\mathcal 
L_u$. If $\mathcal L_u$ does not belong to a {\it cycle}, then by above case, within finite rounds, $r_x$ does not remain non-terminal with the robots on $\mathcal L_u$ and after that $r_x$ will make a type-1 move along $\mathcal L_v$ to break the cycle $\mathcal C$. Again, if $\mathcal L_u$ belongs to a {\it cycle},  $r_x$ is a robot at a critical robot position on $\mathcal L_u$ and  a type-1 movement of $r_x$ along $\mathcal L_u$ breaks this cycle. Thus, within finite time all the cycles in $\widetilde{\mathcal S}$ shall be broken.
 \end{itemize}
Hence, within finite time, $r_i$ becomes a terminal robot. Since robots are semi-synchronous, by lemma-1, $r_i$ remains as terminal once it becomes so.

 \begin{lemma}
\label{lemma6}
The movements of the robots are collision free. 
\end{lemma}

\textbf{Proof.} Let $r_i$ and $r_j$ be two arbitrary robots and at least one of them move. Consider a robot $r_k$ visible to at least one of $r_i$ and $r_j$.  
 If $r_i$ and $r_j$ collide, then $r_i$, $r_j$ and $r_k$ would become collinear or remain collinear which are contradictions to lemma $\ref{non-collinear}$ and $\ref{non-terminal}$.
 This implies that the movements of the robots are collision free during the whole execution of $MutualVisibility()$.
 \begin{lemma}
\label{lemma7}
If $\mathcal R(t_0)\notin \widetilde{C}_L$, during the whole execution of algorithm $MutualVisibility()$, the convex hull of the robot positions in $\mathcal R(t_0)$ remains invariant in size and shape.
\end{lemma}
 \textbf{Proof.} Let $\mathcal {CH}(t_0)$ denote the convex hull of $\mathcal R(t_0)$. The robots occupying the vertices of $\mathcal {CH}(t_0)$ are terminal robots. According to  algorithm $MutualVisibility()$, these robots do not move. Again, the robots on the edges of $\mathcal {CH}(t_0)$
  move inside the convex hull $\mathcal {CH}(t_0)$ and no robot, lying inside the hull, crosses any edge of the convex hull (according to the definitions of directions of movements and amount of displacement in case-1 of subsection D). Hence, $\mathcal {CH}(t_0)$  remains invariant in size and shape.

From the above results, we can state the following theorem:

\begin{theorem}
 Algorithm $MutualVisibility()$ solves the mutual visibility problem without any collision for a set of semi-synchronous, communication-less robots, placed in distinct location, with 1 bit of persistent memory.
\end{theorem}

\section{Conclusion}
\label{con}  
This paper presents a distributed algorithm to solve the mutual visibility problem in
finite time for a set of communication-less semi-synchronous robots endowed with constant amount of persistent memory. The proposed algorithm uses only 1 bit of persistent memory. The  robots use their persistent memories only to remember information about their last movements. There is no explicit communication between the robots. The algorithm also guarantees collision free movements for the robots. The results of this paper leave many open questions. How does the internal persistent memory can help to reduce the communication overheads in the existing solutions for the mutual visibility problem, where external lights are used  for communicating the internal states of the robots? How to solve the mutual visibility problem for asynchronous robots in this setting? What would be the impact of internal persistent memory in the solutions of other geometric problems?

\end{document}